# Safe Use of Neural Networks

G. Robert Redinbo

**Abstract**— Neural networks in modern communication systems can be susceptible to internal numerical errors that can drastically effect decision results. Such structures are composed of many sections each of which generally contain weighting operations and activation function evaluations. The safe use comes from methods employing number-based codes that can detect arithmetic errors in the network's processing steps. Each set of operations generates parity values dictated by a code in two ways. One set of parities is obtained from a section's outputs while a second comparable set is developed directly from the original inputs. The parity values protecting the activation functions involve a Taylor series approximation to the activation functions. We focus on using long numerically-based convolutional codes because of the large size of data sets. The codes are based on DFT kernels and there are many design options available. MatLab simulations show our error-detecting techniques are effective and efficient.

**Index Terms**—Neural networks, convolutional codes, error detection, soft errors, matrix operations, activation functions, DFT-based convolutional codes, algebraic-based fault tolerance (ABFT)

## 1 Introduction

Communication systems can use neural networks in many parts as is outlined in an article describing many applications of neural networks [1]. Neural networks have many processing operaions that are susceptible to random internal numerical errors that can drastically alter their decision outputs. Safe use needs to know when errors have appeared. Networks in commercial situations on standard computing hardware are extremely reliable.

On the other hand, there are situations involving neural networks that operate in what we term, hostile environments, where radiation particles can disrupt normal numerical operations causing erroneous decisions and improper tuning. For example, remote sensing in earth orbit or on foreign planets face disruptions. Neural networks can be used in orbital control systems or in medical systems within high-energy environments. Control systems in heavy industrial locations can be influenced dramatically. This paper addresses a standard neural network configuration and proposes protective methods that can detect when errors have affected numerical calculations voiding safe use.

Neural networks appear in many forms. They all have some common operations in stages forming the network, both in a forward direction when yielding decision results and in a backward propagation needed for tuning and training the network [Chapter 5, 2]. Each stage in the network whether a forward or backward type involves weighting data, scale by coefficients and summing, or passing through activation functions, nonlinear operations with limited output range. We propose numerically based error-detecting codes wherein the code word symbols are numerical values with coding operations defined on an arithmetic field.

The purpose of this paper is to guarantee safe us by detecting errors in operations supporting neural networks. Error-detecting codes are applied to generic models of the stages in neural networks, not aimed for any specific implementation. In this way, these new methods can be appropriately modified to address any practical neural network implementation

The concerns about errors in neural networks have been expressed in many papers with quite different viewpoints and sometimes with new approaches for increasing the protection levels. There are many, many articles concerning neural networks in the literature and some of them address reliablitiy issues. We mention some in particular that seem in the direction of results in this article.

Some papers [3-5] evaluate various architectures that mitigate the effects of single-event upset (SEU) errors in networks of various kinds. A fault-tolerant method in [6] employs training algorithms combined with error-correcting codes to separate the decision result allowing errors to be more noticeable. The training procedures for this approach are complicated. A series of papers [7-10] make hardware changes to the underlying implementation devices to avoid errors. Paper [10] diversifies the decisions steps allowing better detection of errors. When memistrors are employed [11-13] several clustering techniques including binary error-correcting codes on the network's outputs offer some protection.

Several papers are concerned with the impact of SEU errors on the memory system supporting the network [14-16]. One approach focuses on storage of the weight values [16]. Another article addresses modifying the activation func-

Robert Redinbo is with *the Department of Electrical and Computer Engineering, University of California, One Shields Avenue, Davis, CA 95616 USA (e-mail: redinbo@ece.ucdavis.edu.)*

tion (models of neurons) so that failures during these evaluations can be checked more easily.

After we had completed our developments of error-detection codes for protecting the weighting operations and activation function outputs, we discovered a short conference paper (2pages) [18] that started in the same direction as our approach. We recognized their approach as using algorithm-based fault tolerance (ABFT) methods [19]. Their results relied on BCH codes upgraded to real number block codes, which they applied to a three-stage network for simulation purposes. We believe our new results employing numerically based convolutional codes for both weighting actions and activation functions provide a much broader scope for protection techniques.

The next section describes the general features of most neural networks. The following section explains our novel technique for detecting numerical errors in both forward and backward stages aligning with the data flow of the network. The use of wavelet codes, convolutional codes with numbers as symbols and parities, permit the large size of the data vectors to be handled offering protection through and across the weighting and activation function computations. A special method is developed for protecting the calculations producing the new weighting matrices that are determined in the backward stages. Thus, the infrequent tuning of the network as implemented by the backward sections are protected. The last section evaluates the effectiveness of these new codes, showing unusually broad detecting performances. A short appendix outlines the design details of the new wavelet codes.

## 2 Modeling Neural Networks for Protection Evaluation

There are many different neural network configurations for implementing artificial intelligence operations. Our goal is to demonstrate methods for detecting any errors that occur in arithmetic operations in the network. Computational errors arise from the underlying electronic structures' failures such as by soft errors [3]. They are very infrequent and their sources are difficult to pinpoint but can have a devastating effect on the results of a neural network. Accordingly, we will adopt a reasonable model that considers all normal operations that appear in neural networks. We employ a model offered by a text on neural networks [Chapter 5, 2]. The common arithmetic processing features involve large arithmetic weighting operations and activation functions. All are included in Fig. 1, which depicts the weighting sections $W^{(p)}$ and activation operator A.

The data processed in stages are collected into vectors $\underline{Y}$, dimension K, as will be more formally described shortly. The number of data samples passed between stages can be different sometimes, but for exposition purposes, we assume the same size through the stages. The weighting functions are implemented by a matrix W with its scaling features. The outputs of the weighting operations are denoted by $\underline{S}$ whereas the outputs to the next stage are from the activation functions, each of which have a limited output range. The role of the forward stages is to produce decision variables outputs in $\underline{Y}^{(M)}$. These final outputs of the forward network yield the decision variables.

The neural network is adjusted, possibly infrequently, by the actions of the backward propagation stages that process errors between the actual outputs and the desired outputs of the forward stages. These error values are passed through the backward stages. Each backward stage processes errors from the previous stage. Then, based on the perceived errors using the outputs of a comparable indexed forward stage, a new set of weights are computed in a backward stage. These new weights will be employed in a future forward stages. In addition, the newly defined weights are used to continue through the backward propagation processing stages. This approach also detects any errors in the control operations since they ultimately lead to numerical errors.

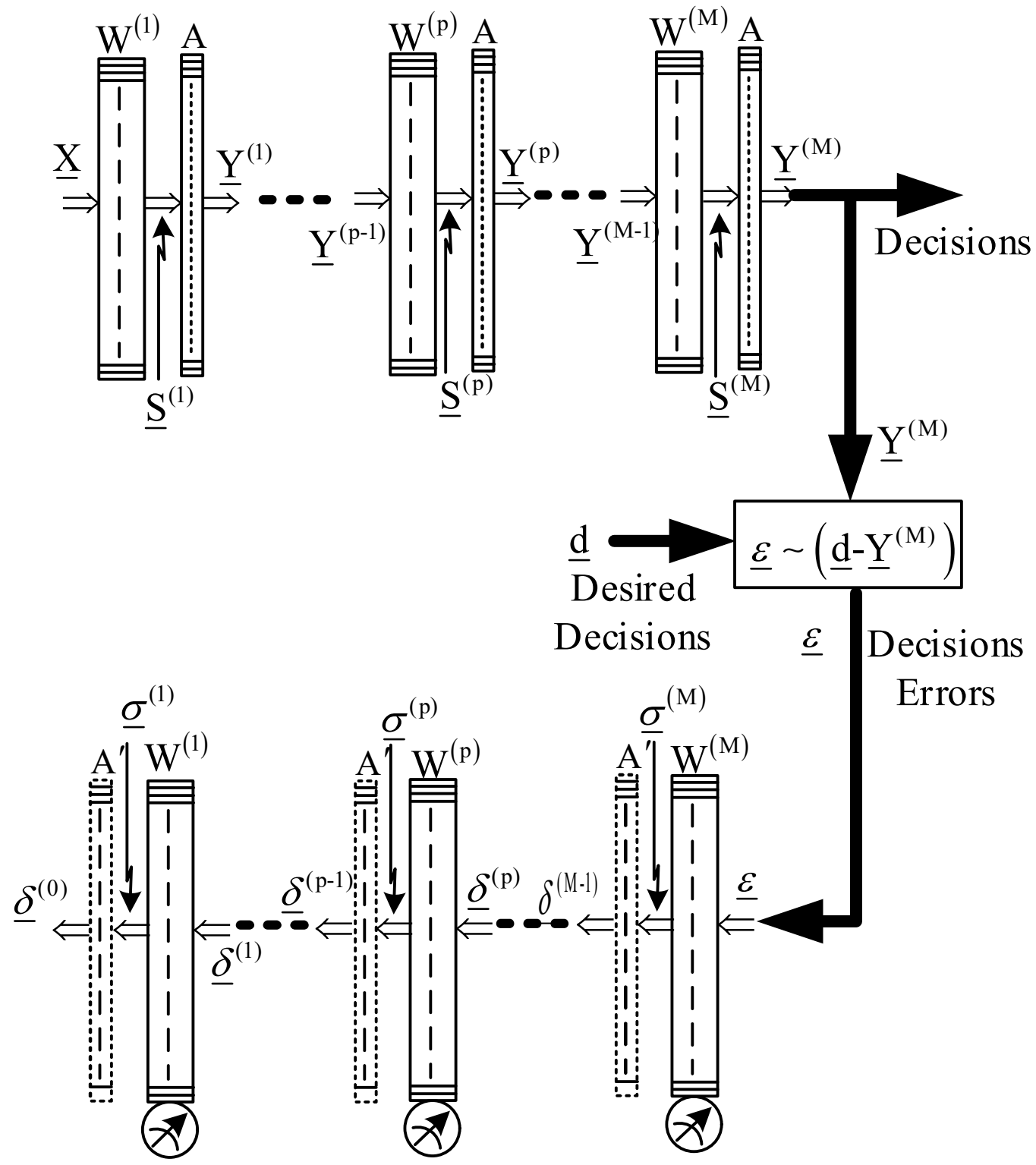

Figure 1. Arithmetic Stages of Neural Network

The arithmetic operations in a typical forward stage, labeled by index p, out of M forward stages is detailed further in Fig. 2. We consider the arithmetic operations as separate entities whether by fixed-point or floating-point nature. Thus, arithmetic values are the fundamental quantities in this work. The inputs to this stage p are outputs of activation functions from the previous stage. These inputs are collected into a vector $\underline{Y}^{(p-1)}$, $1\times K$. They are combined with the weighting matrix $W^{(p)}, K\times K$, yielding outputs $\underline{S}^{(p)}, 1\times K$. The activation function g(x) is applied to each variable in $\underline{S}^{(p)}$ providing the outputs of stage p, $\underline{Y}^{(p)}$. Since the weightings are applied linearly, the processing from $\underline{Y}^{(p)}$ to $\underline{S}^{(p)}$ is a matrix-vector equation.

$$\underline{S}^{(p)} = \underline{Y}^{(p\text{-}1)} W^{(p)} \quad (1)$$

In a compressed notation, the outputs $\underline{Y}^{(p)}$ are expressed by applying g(x), activation function, to each component of $\underline{S}^{(p)}$.

$$\underline{Y}^{(p)} = g\left(\underline{S}^{(p)}\right) \quad \text{Activation Function Outputs} \quad (2)$$

The activation function can take several nonlinear forms, e.g. tanh(x) ReLU(x), [20]. It is useful to have a good derivative since the backward stages use derivatives of the forward function. (Some other functions besides tanh(x) may not have derivatives at all places but that can be handled [20])

The forward stages produce decision variables that give the choices for the output. If the forward network is to be trained, adjusted, the decision variables are compared to the desired outputs and any discrepancies are used by the backward propagation stages to adjust the weighting values in $W^{(p)}$'s. Any future adjusted network will use the new values contained in the new weighting matrices $W^{(p)nw}$. The role of the backward propagation stages is indicated in Fig. 1 by the adjustable symbol under backward stages. Keep in mind that the activation function in the backward section is the derivative of the one in the correspondingly indexed forward stage.

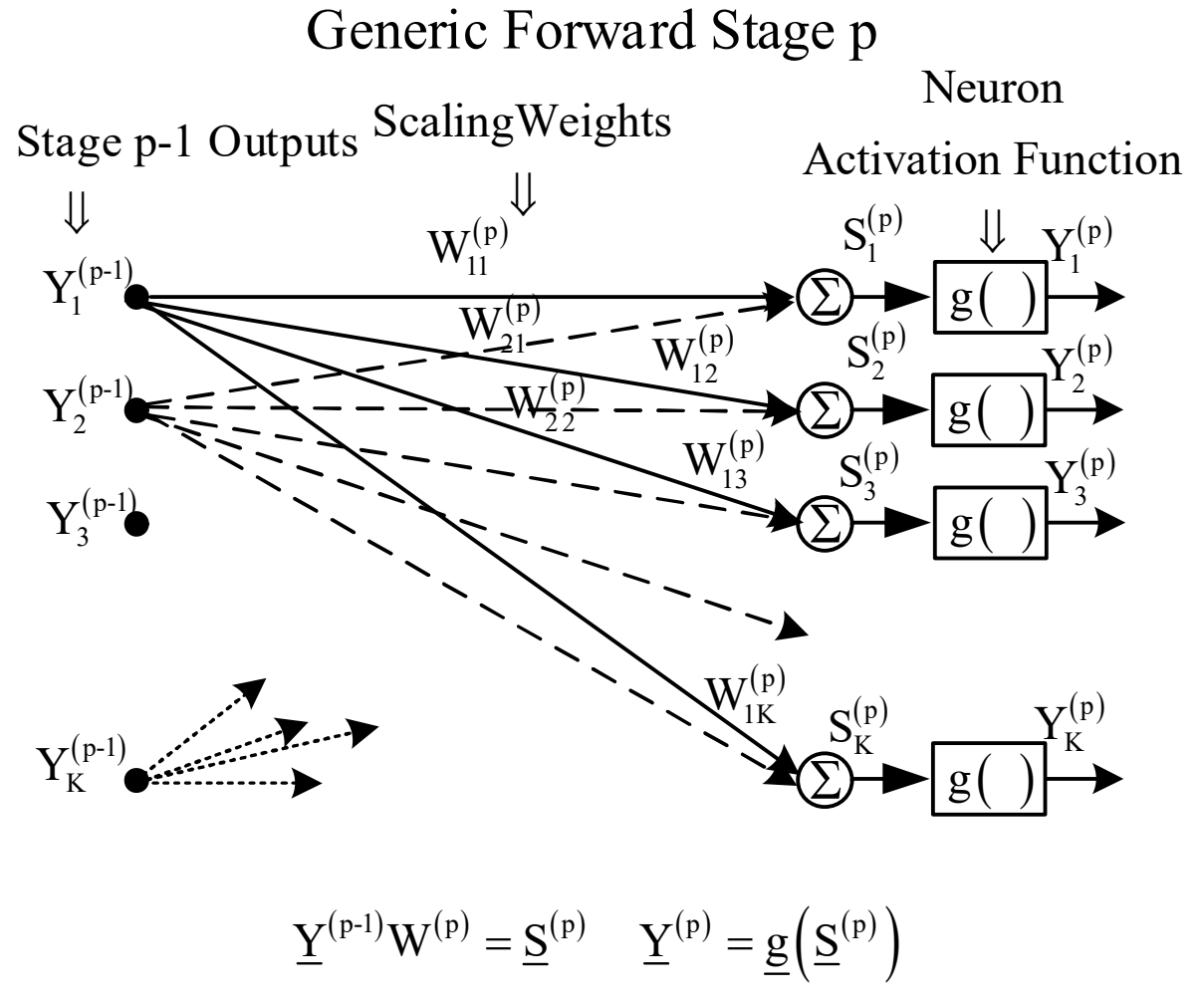

Figure 2. Forward Stage of Neural Network

The update of each weighting matrix is done in two phases. The output of a backward stage, $\underline{\delta}^{(p)}$, as indicated in Fig. 3 showing a generic backward stage p is derived using new weights called $W^{(p)nw}$. This is to designate the new variables in backward stages Fig. 3. They are derived using the propagating error vector $\underline{\delta}^{(p\text{-}1)}$ from the previous backward stage (p+1) and the input vector $\underline{Y}^{(p\text{-}1)}$ saved from the comparably indexed FORWARD stage.

$$W_{ji}^{(p)nw} = W_{ji}^{(p)} + \eta \delta_j^{(p+1)} Y_i^{(p\text{-}1)} \quad (3)$$

$;\eta$ learning rate, $Y_i^{(p\text{-}1)}$forward network values

The output vector $\underline{\delta}^{(p\text{-}1)}$ is calculated using the new weights. Remember, the backward propagation stages are not used during normal decision operations.

## 3 Protecting Processing Operations

There are two major processing operations involved in every forward and backward stage (Fig. 1). The filter weighting calculations may be viewed as a large matrix structure, far larger than most matrix-vector products. The second part of each stage has a bank of activation functions. It might seem that the backward feedback which adjusts the filter weightings could eventually correct any errors introduced in the updating of the weights. However, it is easy to conjure up situations where any errors would continue through to the feedback. We propose using error-detecting codes defined over large arithmetic fields to determine when computational errors in either part of a stage are present. When errors are detected, the processing steps need to be repeated to support safe use.

### 3.1 Weighting Operations

The weighting of input data in stages is effectively a vector-matrix operation. For example, the data input to stage p of forward section, Fig. 2, weights a vector $\underline{Y}$, $1\times K$, with matrix W, $K\times K$. The output values are placed in a K vector $\underline{S}$.

$$\underline{S} = \underline{Y}\, W \qquad 1\times K \text{ vector} \quad (4)$$

There can be occasional infrequent sparse numerical errors in the resulting vector caused by failures in the vector-multiply operations. (Later, we will model these errors as additive errors inserted in the components of W.)

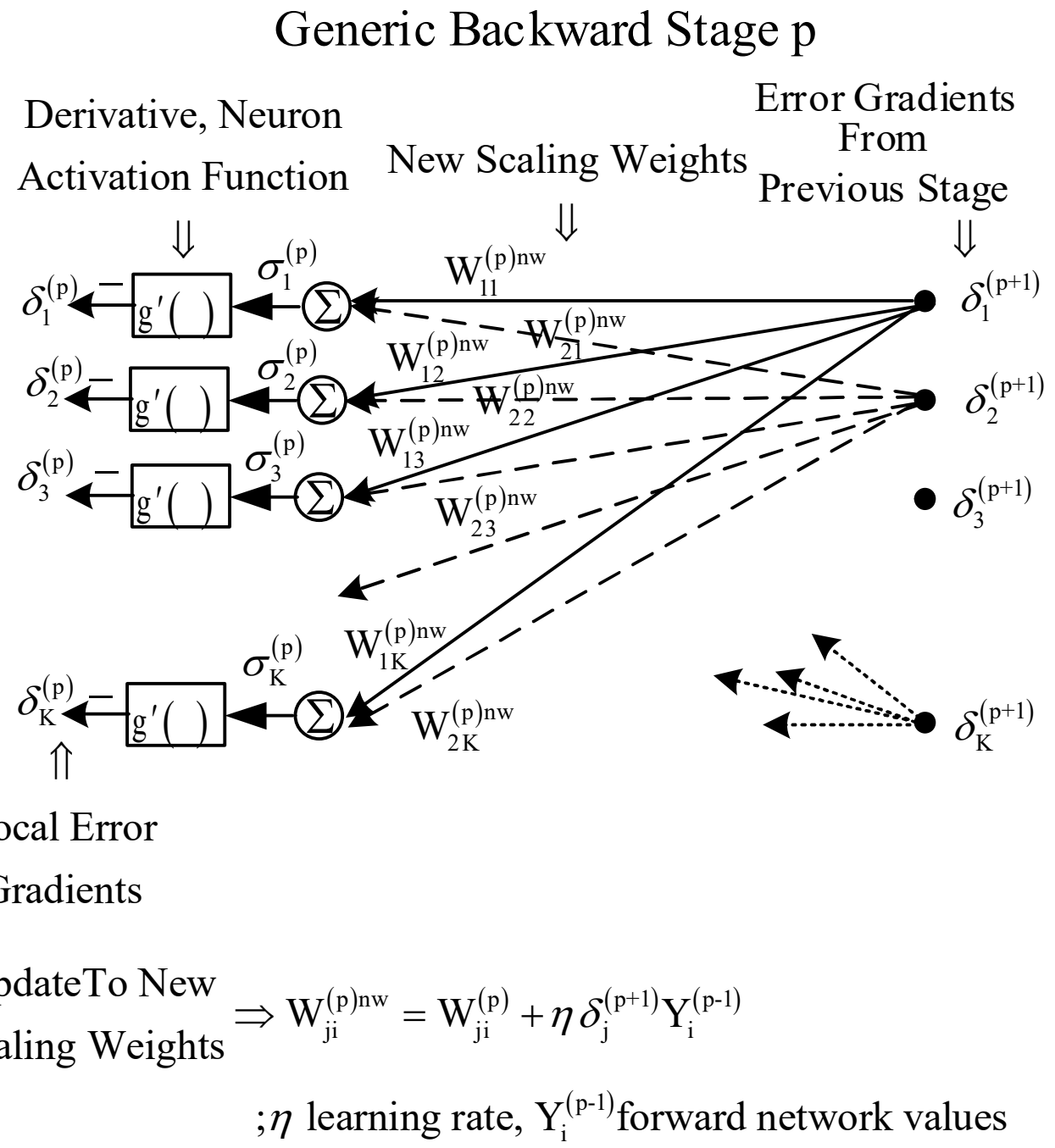

Figure 3. Backward Propagation Stage of Neural Network

Error-detecting codes can be employed to sense any er-rors in $\underline{S}$. We are considering block codes first for describing the concepts. Later, we will expand the approach to employ wavelet codes. Think of a large length code word in systematic form (data and check symbols are distinct), defined by an encoding matrix $G_S = (I_k\ P)$. The parity-generating matrix for a linear code is $P,\ K \times (N\text{-}K)$, for a length N code with K data positions.

We will use a methodology called algorithm-based fault tolerance (ABFT) [19] where the parity values associated with output vector $\underline{S}$ will be computed in two independent ways and then compared. One set of parity values can be developed:

$$\underline{\rho} = \underline{S}P \qquad 1 \times (N\text{-}K) \qquad (5a)$$

However, $\underline{S}$ results from using matrix W applied on the inputs $\underline{Y}$. Thus, an alternative version for the parity values can be defined.

$$\underline{\rho}_a = \underline{Y}\,WP \qquad 1 \times (N\text{-}K) \qquad (5b)$$

When parities in $\underline{\rho}$ and $\underline{\rho}_a$ are compared, if they disagree in even one place, errors appear in $\underline{S}$ or $\underline{Y}$ or both (up to error-detecting capability of code).

The matrix product (WP) appearing in developing $\underline{\rho}_a$ is smaller than W; WP is $K \times (N\text{-}K)$. It can be formed beforehand independently. The overhead in this ABFT scheme is in the calculation $\underline{Y}\,WP$ (5b) and $\underline{S}\,P$ (5a). Note the computation of the output $\underline{Y}$ using W is already required for the forwarding data. The efficiency of the ABFT method relies on the value (N-K) being much smaller than K the data size. The overall concept of ABFT is shown in Fig. 4.

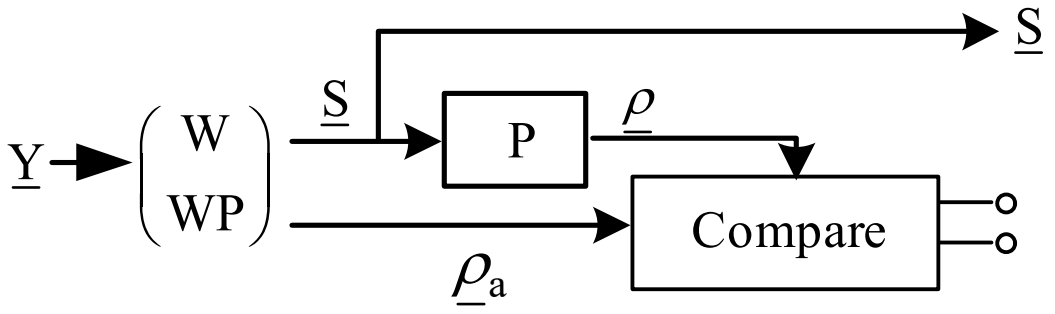

Figure 4. Algorithm-Based Fault Tolerance Method

The comparison of parities in $\underline{\rho}$ and $\underline{\rho}_a$ allows a small tolerance in each position, particularly when no errors are present since roundoff noise can enter the two different calculations. Of course, it is always possible to overwhelm any error-detecting structure by encountering too many errors. This is true for all error-detecting codes. No errors guarantees safe processing.

We can use the same parity-check matrix P to produce parities associated with the calculation of the new weighting function in each stage in the backward direction. A generic backward stage first develops an updated weighting matrix, which is employed in the next forward stages and henceforth until the next training sessions. We can describe a method for verifying proper calculations using a generic backward stage with this new updated weighting called $W^{nw}$ with formula similar to (3).

$$W^{nw} = W + \eta\underline{\delta}^T\underline{Y}; \quad \begin{array}{l} \eta \text{ is learning factor} \\ \delta, 1 \times K \text{ error gradient} \\ \underline{Y},\ 1 \times K \text{ input data forward} \\ W,\ K \times K \text{ matrix forward stage} \end{array} \qquad (6)$$

The parity-check matrix P is applied to $W^{nw}$ generating a row of parity $\rho_i^{nw}$.

$$\left(\left(\rho_i^{nw}\right)\right) = W^{nw}P \quad K \times (N\text{-}K) \qquad (7a)$$

Each of the K rows of $\left(\left(\rho_i^{nw}\right)\right)$ hold (N-K) parity values. However, similar parity values can be computed by applying P to the two parts of (6) individually and adding.

$$\left(\left(\rho_i'\right)\right) = WP + \eta\underline{\delta}^T\underline{Y}P \qquad (7b)$$

Note, two items in this parity equation have been calculate in the forward operation. $\underline{Y}P,\ 1 \times (N\text{-}K)$, was formed in the forward stage as was WP. Then scaling the $\underline{Y}P$ by the error gradient $\underline{\delta}^T$ produces a matrix $K \times (N\text{-}K)$ which when adjusted by $\eta$ and added to WP yields K new row vectors each $1 \times (N\text{-}K)$. When $\left(\left(\rho_i^{nw}\right)\right)$ rows are compared to the rows of $\left(\left(\rho_i'\right)\right)$, any mismatch indicates a detected error in the calculation of $W^{nw}$.

$$\left(\left(\rho_i^{nw}\right)\right) \sim \left(\left(\rho_i'\right)\right) \quad \text{Parity Comparisons} \qquad (7c)$$

Once each updated weighting vector is computed and checked, it is employed in the backward stage. Now checking this backward stage, its output shown generically as vector $\underline{\sigma},\ 1 \times K$, can produce (N-K) parity values using check matrix P.

$$\underline{\rho} = \underline{\sigma}P \qquad (8a)$$

However, as before, the inputs to backward stages can be used to generate another set of comparable parities using the new weighting matrix just computed. This alternate parity vector is designated $\underline{\rho}_a$.

$$\underline{\rho}_a = \underline{\delta}\left(W^{nw}P\right) \quad 1 \times (N\text{-}K) \quad \underline{\delta}, 1 \times K \text{ input vector} \qquad (8b)$$

Fig.5 shows this ABFT method for a generic backward step.

Long error-detecting block codes have very poor performances, regardless whether finite field or numerically based forms. One way to obtain long error-detecting codes over numerical fields is through convolutional codes. We pioneered a new form of convolutional codes over the complex numbers [24]. Their construction uses the popular discrete Fourier transforms (DFT). A brief description with defining equations is contained in Appendix A.

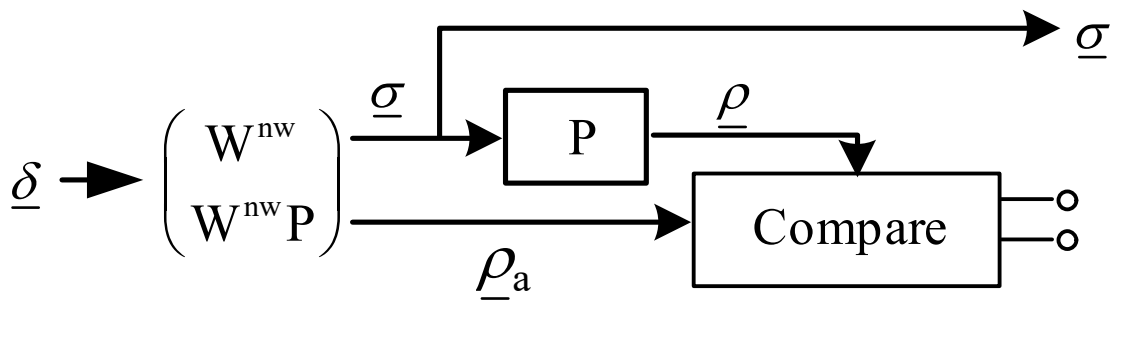

Figure 5. Algorithm-Based Fault Tolerance Backward Propagation Stage

Convolutional codes employ a sliding memory segment of data, linearly producing parity samples. This memory usually called the constraint length of data provides error detection along the codeword stream. We propose such codes to protect the large-sized weighting operations, again employing the concept of algorithm-based fault-tolerance.

The codeword in a standard systematic form intermingles the parity samples among the data segment allowing detecting operations to proceed using a sliding memory. However, it is always possible to handle the parity parts separately. The coding structure we developed actually uses a subcode giving a slightly shorter information parameter. The underlying DFT codes have parameter n length, (k-1) information positions and a constraint length $\mu$ (memory length parameter). There is a constraint among these parameters required.

$$n>(\mu+1)(n\text{-}k);\quad n,k \text{ and } u \text{ , parameter DFT code} \tag{9}$$

We change the notation for the processing and coding to match the traditional form of convolutional codes even though they are viewed here as a "block" code. The data are contained in a column vector as are the associated parity values. If the K data samples are collected in a wide vector $\underline{Y}$ of L, (k-1) subvectors while the affiliated parity values are contained in a vector $\underline{\rho}$ , L(n-k) long

$$\underline{Y}=(\underline{Y}_0,\underline{Y}_1,\ldots,\underline{Y}_{L-1});\ \text{each } \underline{Y}_i\ 1\times(k\text{-}1),\ \underline{Y}\ 1\times L(k\text{-}1),\ K=L(k\text{-}1) \tag{10a}$$

$$\underline{\rho}=\underline{Y}\,\Gamma^T\ ;\ \begin{array}{ll} 1\times L(n\text{-}k+1) & \text{parities} \\ \Gamma^T, L(k\text{-}1)\times L(n\text{-}k+1) & \text{parity-generating matrix}\end{array} \tag{10b}$$

Let L, (k-1) length sub-vectors combine to give a segment of K data samples $\underline{Y}$ (10a). The parity values associated with this segment are determined by a string of $(\mu+1)$ submatrices each $(n\text{-}k+1)\times(k\text{-}1)$ labeled $\Xi_i$.

$$\Xi=(-\Xi_0,-\Xi_1,\ldots,-\Xi_i,\ldots,-\Xi_\mu);\quad (n\text{-}k+1)\times(\mu+1)(k\text{-}1) \tag{11a}$$

This matrix segment is used in proper places inside $\Gamma^T$, as in Appendix A, (A-17). A group of (n-k+1) parity values are generated by applying matrix group (11a) to segment of $(\mu+1)$ subvectors $\underline{Y}_i$ each $1\times(k\text{-}1)$.

$$\underline{\rho}_r=(\underline{Y}_r,\underline{Y}_{r+1},\ldots,\underline{Y}_{r+\mu})\Xi^T \quad ;\underline{Y}_i\ 1\times(k\text{-}1),\ \Xi^T\ (\mu+1)(k\text{-}1)\times(n\text{-}k+1) \tag{11b}$$

The next parity subsector $\underline{\rho}_{r+1}$ encodes the next adjacent group of $(\mu+1)$ subsectors starting with data subvector $\underline{Y}_{r+1}$. Such a code detects (2(n-k+1) additive errors in this system.

The parity scheme can detect errors in the processing operations around matrix W. One set of parities is produced from the output data say in forward stage p, omitting the indices below for brevity. Note, notation employs data and parities in row vector form.

$$\underline{S}=\underline{Y}W\quad ;\ 1\times K\ \text{ stage outputs} \tag{12a}$$

The corresponding output parities are:

$$\rho=\underline{S}\Gamma^T\ ;1\times(n\text{-}k+1)L \tag{12b}$$

An alternate but equivalent set of parities come through the combined parity matrix $(W\Gamma^T)$.

$$\rho_a=\underline{Y}(W\Gamma^T)\quad ;(n\text{-}k+1)L\times 1 \tag{13}$$

The parity checking arrangement comparing two independently computed sets is shown in Fig. 6.

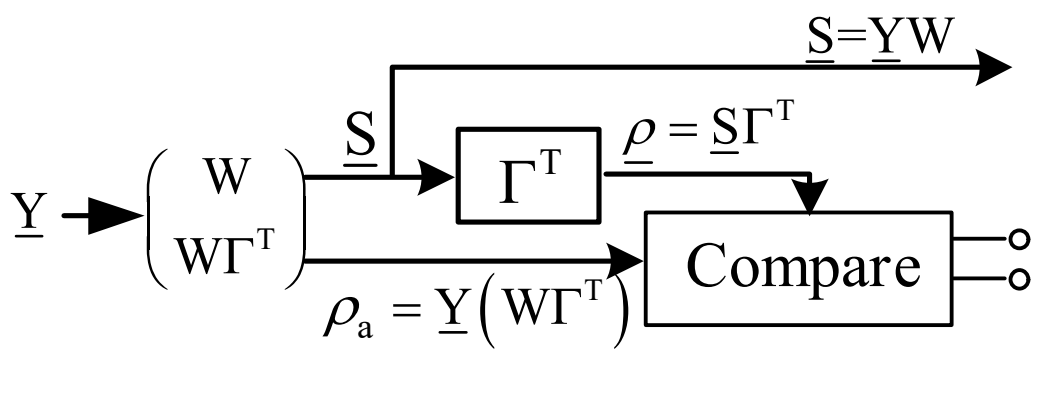

Figure 6. Algorithm-Based Fault Tolerance Using Convolutional Code, Forward Stage

### 3.2 Activation Function Operations

The second step in each stage, both forward and backward directions involves activation functions or derivatives of the activation functions used in the forward paths. There are numerous forms of these functions as collected in a Wikipedia site [20]. We have considered the hyperbolic tangent function, tanh(x), in our developments. As mentioned earlier, the derivatives of activation functions that are needed in the backward stages can be approximated for other types of functions, e.g., ReLU. The forward function is denoted by g(X) and its derivative g′(x), as appear in earlier Figs. 2-3.

The ABFT protection methodology is not automatically applicable to data samples passing through the activation functions because this nonlinear operation acts on each individual sample. However, we are able to present an error-detecting scheme that uses parity generating mechanism similar to the ones in the weighting sections. We remind the reader that we employ data in a vector notation to indicate that all components are handled individually through the activation functions.

$$\underline{Y}=g(\underline{S});\quad \text{Activation Function Processing} \tag{14}$$

A parity generating matrix P, $(N\text{-}K)\times K$, employed earlier for detection techniques produces (N-K) parity values from the vector $\underline{Y}$.

$$\underline{\rho}=\underline{Y}\Gamma^T;\ \text{Parities from Output} \tag{15}$$

It is common to approximate functions even activation functions using a series expansion. A second set of parities can be defined directly from the inputs to the activation function block by examining the Taylor series expansion of the activation function. This is valid if the function is differentiable in the important region of the input $\underline{S}$ items.

$$g(\underline{S})=g(\underline{S})|_{(\underline{S})=\underline{0}}+\frac{dg}{d\underline{S}}|_{(\underline{S})=\underline{0}}\,\underline{S} +\frac{1}{2!}\frac{d^2g}{d\underline{S}^2}|_{(\underline{S})=\underline{0}}\,\underline{S}^2+\cdots+\frac{1}{m!}\frac{d^mg}{d\underline{S}^m}|_{(\underline{S})=\underline{0}}\,\underline{S}^m+\text{error terms} \tag{16}$$

This is an expansion around $\underline{S} = \underline{0}$. Several coefficients in this expansion are known constants and will be designated by elements $A_i$.

$$A_i = \frac{1}{i!}\frac{d^i g}{d\underline{S}^i}\Big|_{(\underline{S})=\underline{0}} \tag{17}$$

Again, using vector notation, the output vector $\underline{Y}=g(\underline{S})$ can be written as an approximation out to (m+1) terms.

$$g(\underline{S}) = \sum_{i=0}^{m} A_i \underline{S}^i \text{ ; where } \underline{S}^i = \left(S_1^i, S_2^i, \ldots, S_K^i\right)^T \tag{18}$$

An alternate set of parities may be determined using the approximation to g(X) and incorporating the parity-generating matrix $\Gamma^T$.

$$\underline{\rho}_a = g(\underline{S})\Gamma^T = \sum_{i=0}^{m}\left(A_i\Gamma^T\right)\underline{S}^i \tag{19}$$

The linearity of the expansion permits the parity matrix $\Gamma^T$ to be applied to individual powers. The parity matrix guarantees that the codewords have good separation insuring detection of errors up to a certain level [25]. Nevertheless, Fig. 7 outlines the ABFT protection scheme using the Taylor series expansion.

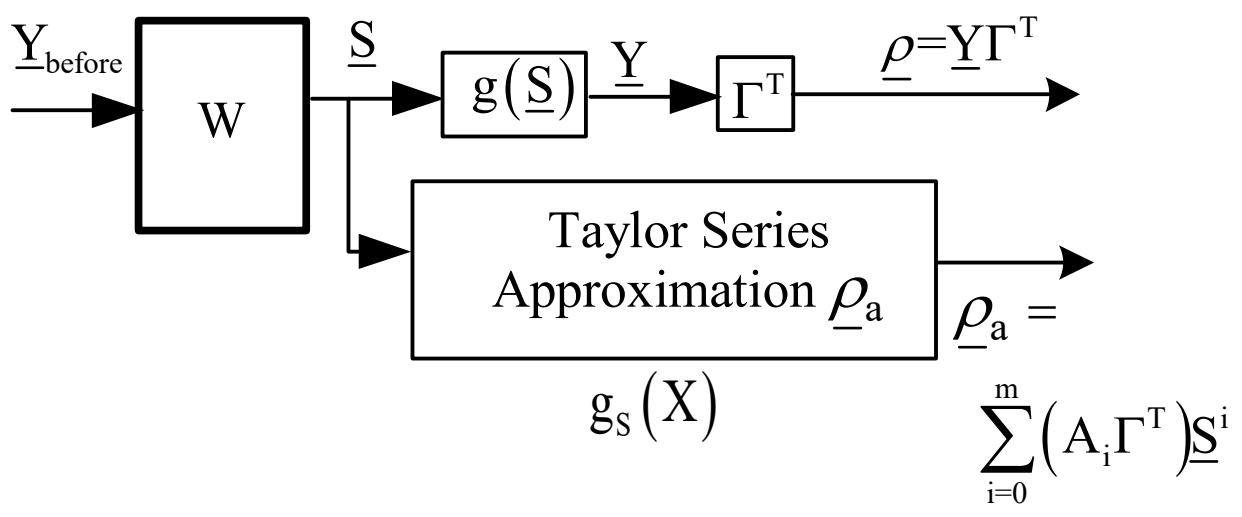

Figure 7. Parity Detection Method Using Taylor Series Expansion of Activation Function

Of course, the powers of items in vector $\underline{S}$ have to be computed and the expansion in (16) may be the basis for computing the activation function itself.

The Taylor series expansion (18) is only valid over a limited range of the components of $\underline{S}$. So, extending the range to develop the alternate parity calculations (19) requires approximating tanh(X) outside |X|<1. The Taylor series for tanh(X) of with up to 10 terms is called g(X).

$$g(x) = x - \left(\frac{1}{3}\right)x^3 + \left(\frac{2}{15}\right)x^5 - \left(\frac{17}{315}\right)x^7 + \left(\frac{63}{2835}\right)x^9 \tag{20}$$

On way to do this extension approximation is to assign fixed values over segments outside the good range. A reasonable approximation for tanh(X) is given $g_S(x)$.

$$g_S(x) = \begin{cases} g(x) & |x| \le 1 \\ 0.8049 & 1 < |x| \le 1.25 \\ 0.8767 & 1.25 < |x| \le 1.5 \\ 0.9233 & 1.5 < |x| \le 1.75 \\ 0.9527 & 1.75 < |x| \le 2 \\ 0.9820 & 2 < |x| \le 5 \\ 1 & 5 < |x| \end{cases} \tag{21}$$

The fixed segments of range of X where constant values are defined is shown in Fig. 8. The constant values are chosen as the midpoint value between endpoints of the segment. The approximation error introduced by these segment values are on the order of $10^{-2}$. The approximation errors are reflected into the parity calculations when $g_S(x)$ is substituted in (19).

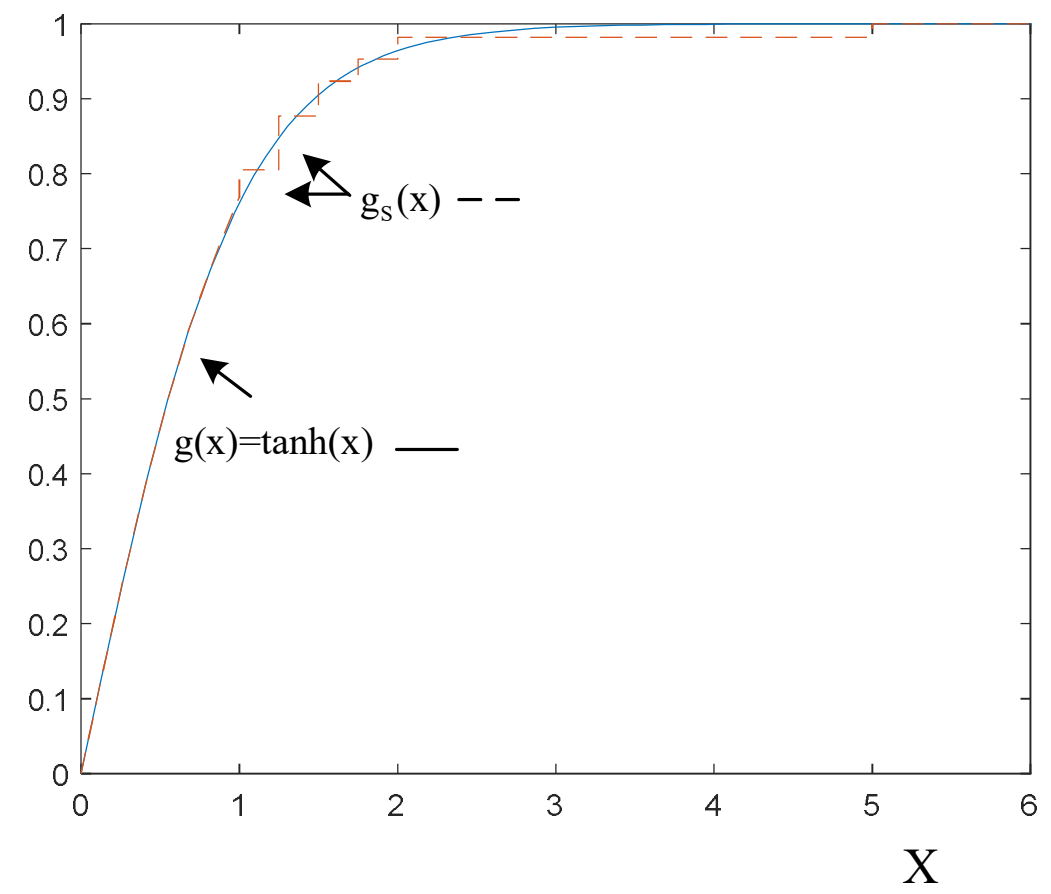

Figure 8. Approximations of tanh(X) by Taylor Series and Constant Value Segments

Similar protection methods can be applied for the derivative of the activation function appearing in the forward sections. The derivative of tanh(X) is $\tanh'(X) = \left[1 - \tanh^2(x)\right]$ which has a diminishing curve from 1 to 0 for positive values of X. Also, the Taylor series for $g'(X)$ is extracted for the expansion for g(X) by taking derivatives inside (20).

$$g'(\underline{S}) = + \frac{dg}{d\underline{S}}\Big|_{(\underline{S})=\underline{0}} \underline{S} + \frac{1}{2!}\frac{d^2 g}{d\underline{S}^2}\Big|_{(\underline{S})=\underline{0}} \underline{S}^2 + \cdots + \frac{1}{(m+1)!}\frac{d^{m+1}g}{d\underline{S}^{m+1}}\Big|_{(\underline{S})=\underline{0}} \underline{S}^m + \text{error terms} \tag{22}$$

Then an approximation using segmented pieces like $g_S(x)$ in (21) is easily established for protection purposes.

On the other hand, if the evaluation functions are implemented by look-up tables, a simple straightforward and effective error-detecting scheme uses duplication. Two compatible outputs from the functions employing two different look-up operations per position. This simple but brute force detection philosophy is outlined in Fig. 9.

## 4 EVALUATIONS

Will the protection methods work, how well and what are the overhead processing costs? Two parts of the neural network stages are addressed by our detection techniques. The large weighting operations, viewed as matrix multiplication, are protected by long numerical-based error-detecting codes. In addition, the bank of activation functions can be protected by similar codes using a segmented Taylor series approximation for parity generation.

The large filtering operations are efficiently covered by DFT motivated codes. An algorithm-based fault tolerance (ABFT) methodology produces two sets of comparable parity values related to the output data weighted by matrix W. Equations (12) and (13) outline this approach relying on a parity-generating matrix $\Gamma^T$. (See Fig. 6) The matrix W, $L(k\text{-}1)\times L(k\text{-}1)$, gives an output $\underline{S}$ that in turn yields a set of Ln_k =L(n-k+1) parity values. The extra cost of the parity generation is a matrix-vector multiplication $\underline{S}\Gamma^T$. Using the number of multiplications to form the parity components in $\underline{\rho}$ (5a) as an indication of extra cost gives $\text{Ln_k}\left(L(k\text{-}1)\right)^2$ multiplications. ($\Gamma^T$ (A-17) contains only short spans of nonzero pieces, which could greatly reduce this number.)

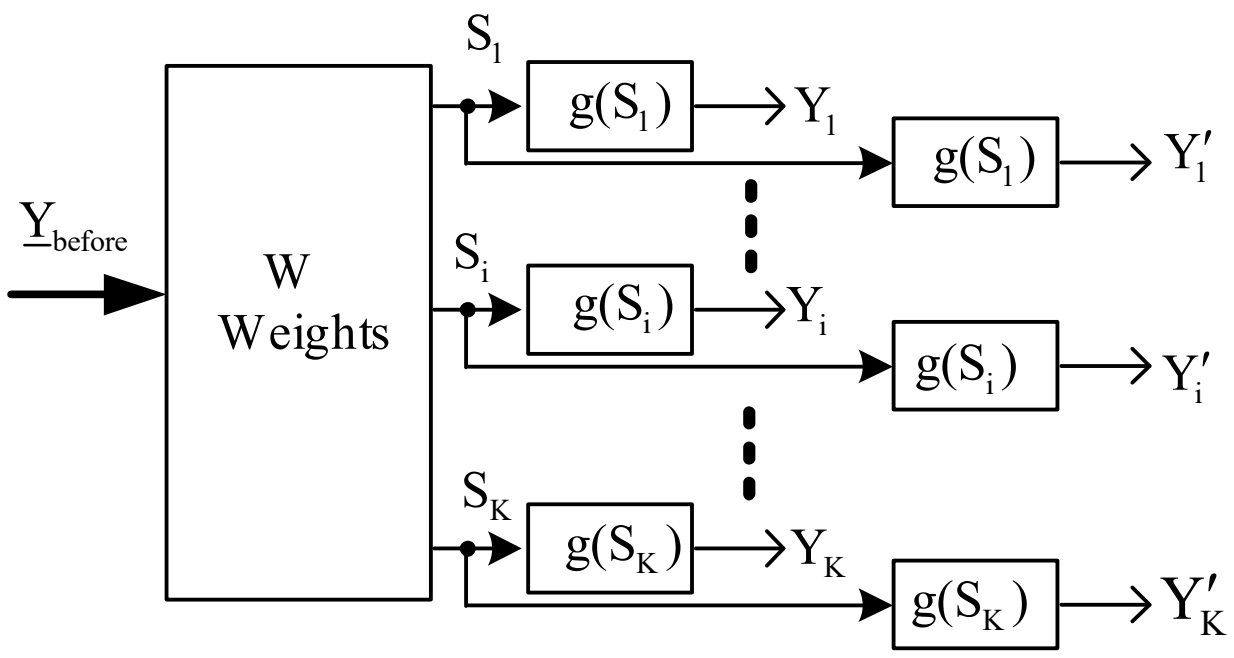

Figure 9. Protecting Activation Steps by Duplication

The other set of parities in $\underline{\rho}_a$ (13) results from applying the combined parity matrix $\left(W\Gamma^T\right)$ which is smaller at $L(n\text{-}k\text{+}1)\times L(k\text{-}1)$. This matrix can be computed off-line (not required at data processing steps). It operates on the input data in $\underline{Y}$ to matrix W. This second set of parities also needs products. $L(n\text{-}k\text{+}1)\left(L(k\text{-}1)\right)^2$. Thus, the parity calculations involve $2L(n\text{-}k\text{+}1)\left(L(k\text{-}1)\right)^2$ extra multiplications. This is the operational overhead required to insert error-detecting codes in the stages.

We ran extensive MatLab simulations concerning the processing of data with weighting matrix W. These simulations focus on errors inserted in the numerical values during processing. The experiments considered three instances of matrix processing, one modeling the forward section and two backward sections. One of the background steps used the same matrix from a similarly indexed forward section while the other backward propagation section used the new weighting matrix $W^{nw}$ (6), resulting from the updating actions of the forward matrix W.

The simulations go through three executions of the matrix iterations with randomly generated input data for each pass. This guarantees virtually everything is random to test all operations. The total number of passes numbered in the millions. The errors in multiplications in the processing steps are modeled by randomly selecting locations in the matrix W into which random sized errors are added. The additive error values follow a Gaussian-Bernoulli probability law. The positions where these random errors are inserted are selected uniformly distributed over all positions in the matrix with independent probability $\varepsilon$. Fig. 10 indicates the main loop for one pass of the simulation.

The operations are protected by a DFT-based convolutional code (wavelet code in Appendix A). The parameters of each code are basic length of pieces n, (k-1) information content, and number of parity values per code piece (n-k+1) with constraint length $\mu$. The two parity sets for each section of a pass are compared and if any position differs between the two, an error is detected. (Many positions of parity could deviate, still counting as a detection.) The independent insertion probability $\varepsilon$ was varied over 13 values ranging from $\varepsilon = 10^{-3}$ to $\varepsilon = 0.1$ and during each value, $10^7$ passes were executed. The number of errors detected were collected for each pass. The performance of the coding was so good that no errors were missed. Thus, the probability of detection for runs in these ranges was 1.0

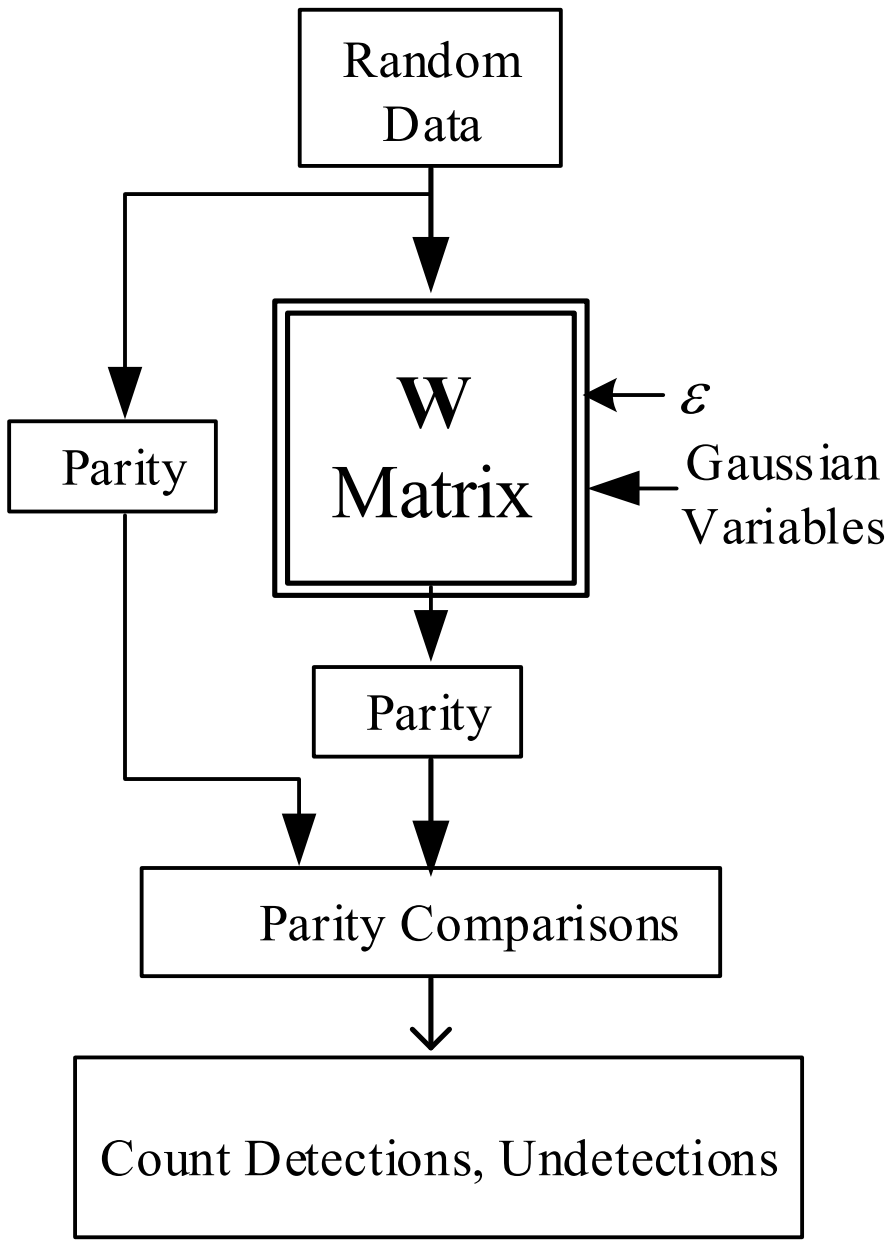

Figure 10. Basic Simulation Loop, Filter Section

The simulations were performed for four different numerical-valued convolutional codes whose different parameters are given in Table 1. The sizes of the matrix W used in each suite of simulations depended on a parameter L=12 and the parameters k and n-k. The matrix W was $L(k\text{-}1)\times L(k\text{-}1)$ and the number of parity values used in each case was L(n-k+1).

The number of errors that were detected were counted. For each data point, 4 million passes were made. The codes are intrinsically powerful so that only a few errors were missed (under 5 per pass). However, if the insertion probability $\varepsilon$ is raised above 0.5, errors were missed because they exceeded the capabilities of the codes. The number of errors detected for each insertion probability $\varepsilon$ and for each of the codes employed is plotted in Fig.11.

Table 1
Simulation Code Parameters

| Code | k-1 | (n-k+1) | $\mu$ | L(k-1) | L(n-k+1) |
|---|---|---|---|---|---|
| 1 | 8 | 4 | 2 | 96 | 48 |
| 2 | 11 | 4 | 2 | 132 | 48 |
| 3 | 11 | 5 | 2 | 132 | 60 |
| 4 | 10 | 3 | 10 | 120 | 36 |

The evaluation of the activation function processing protection scheme is much simpler than that of the weighting sections. Of course, a protection method by duplicating the bank of functions is very straightforward and so we did not simulate such a case. We assume for our purposes that the activation functions and their derivatives were based on tanh, the hyperbolic tangent and its derivative. We simulated the protection method that employed the approximation technique of a modified Taylor series. The outputs from normal processing using tanh(x) activation are compared to those from the Taylor series approximation of activation function Fig. 12.

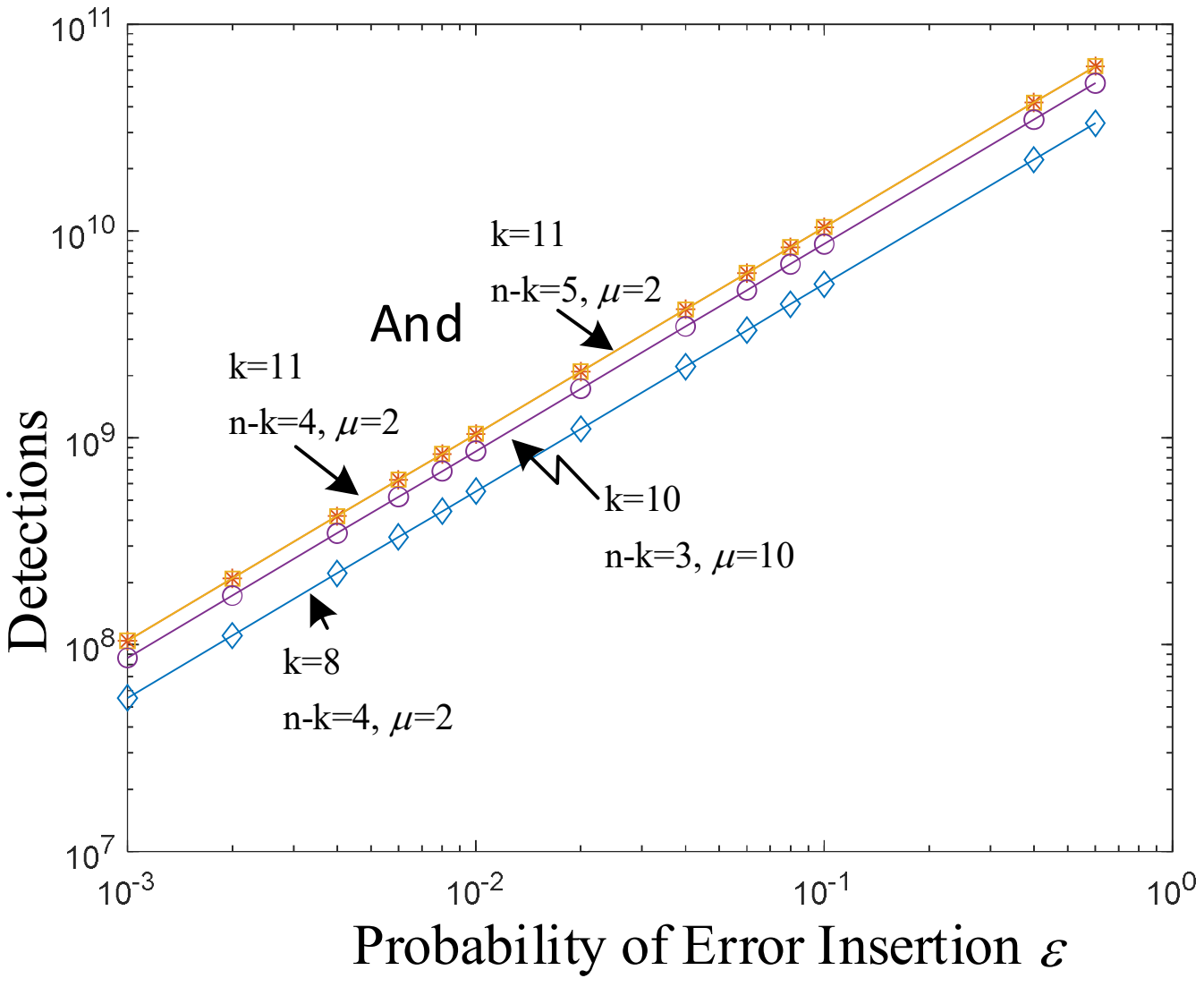

Figure 11. Detection Values of Protected Matrix Operations

For the case of activation function processing, simulations used inserting errors on top of the normal outputs. The first parity was directly from these outputs, possibly corrupted with infrequent additive errors. The other parity set of values used the tanh() approximation in (21), Fig. 8. A comparison of these parity values provide detection capabilities. However, this approximation increased the level of small errors appearing in this set of parities. The Taylor series part utilizes powers of input samples. It is hard to estimate its overhead because of the randomness of samples.

The look-up part of the segmented Taylor series-based approximation introduces errors when normal error-free data are processed through it. Thus, the threshold for determining if parities are mismatched has to be increased. Nevertheless, the segmented part could cause ( infrequently) miss-detections. These, in turn, can prompt some processing to be repeated after incorrectly detecting errors in this part. We simulated millions of passes with no errors present and quickly found that there was a discernable threshold above which normal error-free data processing caused no false detections. The threshold was increased by a factor of 10 over that used in the weighting simulations.

Once, we had set a threshold properly to avoid false errors being detected, simulated errors were properly detected all the time. We ran long simulations inputting random data and selecting error positions in the output places to add Gaussian errors. The results were gathered as before, and detection levels are plotted in Fig. 13. We note the number of detected errors are of course much less since there are only L(k-1) places where errors are inserted compared to the case of weighting matrix W that had $(L(k\text{-}1))^2$ places for error insertions. The detection performances, when approximations for the activation function are employed, are shown in Fig. 13.

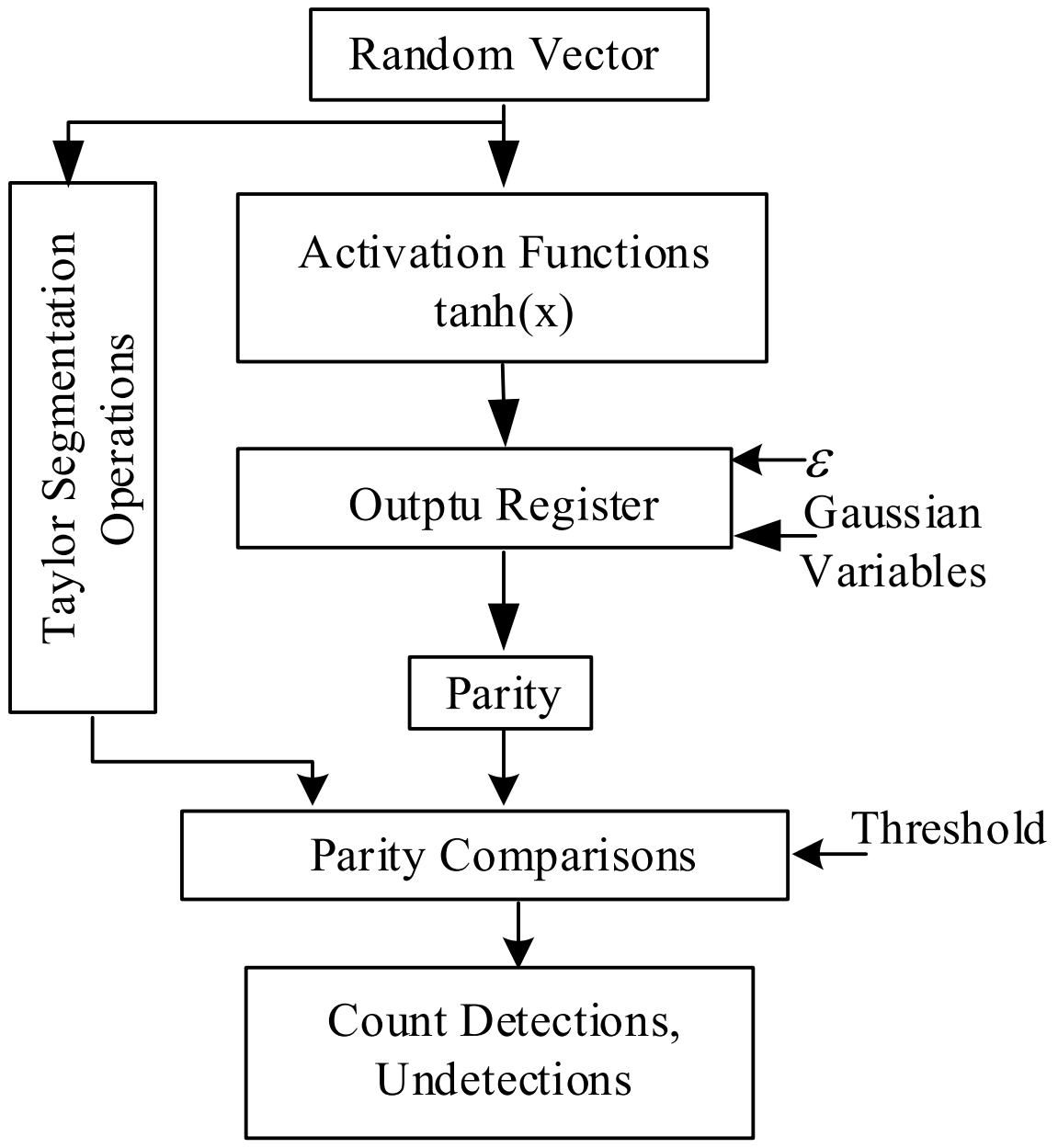

Figre 12. Simulation Activation Functions Section

## 5 Summary

Neural networks employ several stages that each include weighting operations and activation functions. Safe operation requires no errors appearing in any stage. The weighting parts involve a large number of arithmetic operations, multiplications and additions, which could be susceptible to infrequent numerical errors. The activation functions in the stages, which are nonlinear and have limiting outputs, can also suffer numerical errors. In all cases, errors can drastically change the final decisions at the outputs of the network.

This paper proposes and evaluates methods using numerically based error-detecting codes to sense any processing errors that can led to erroneous decisions. The elements in the code are arithmetic symbols, not bits or groups of bits as used in more familiar error-detecting codes. We explored long codes in both sections of a network stage, weighting and activation functions. The detecting procedures involve generating parity values dictated by the code in two ways; one set from the outputs and the other directly from inputs to the process stage. Protecting activation functions' operations uses a segmented Taylor series to generate one set of the necessary parity pairs. We also showed a technique for detecting numerical errors in computing the updated weighting functions implemented in the backward stages.

Extensive MatLab simulations modeling errors in a hostile environment of both weighting and activation function operations show this approach is extremely effective in detecting randomly occurring errors in the processing steps. The Taylor series approximation method for generating checking parities gives good results.

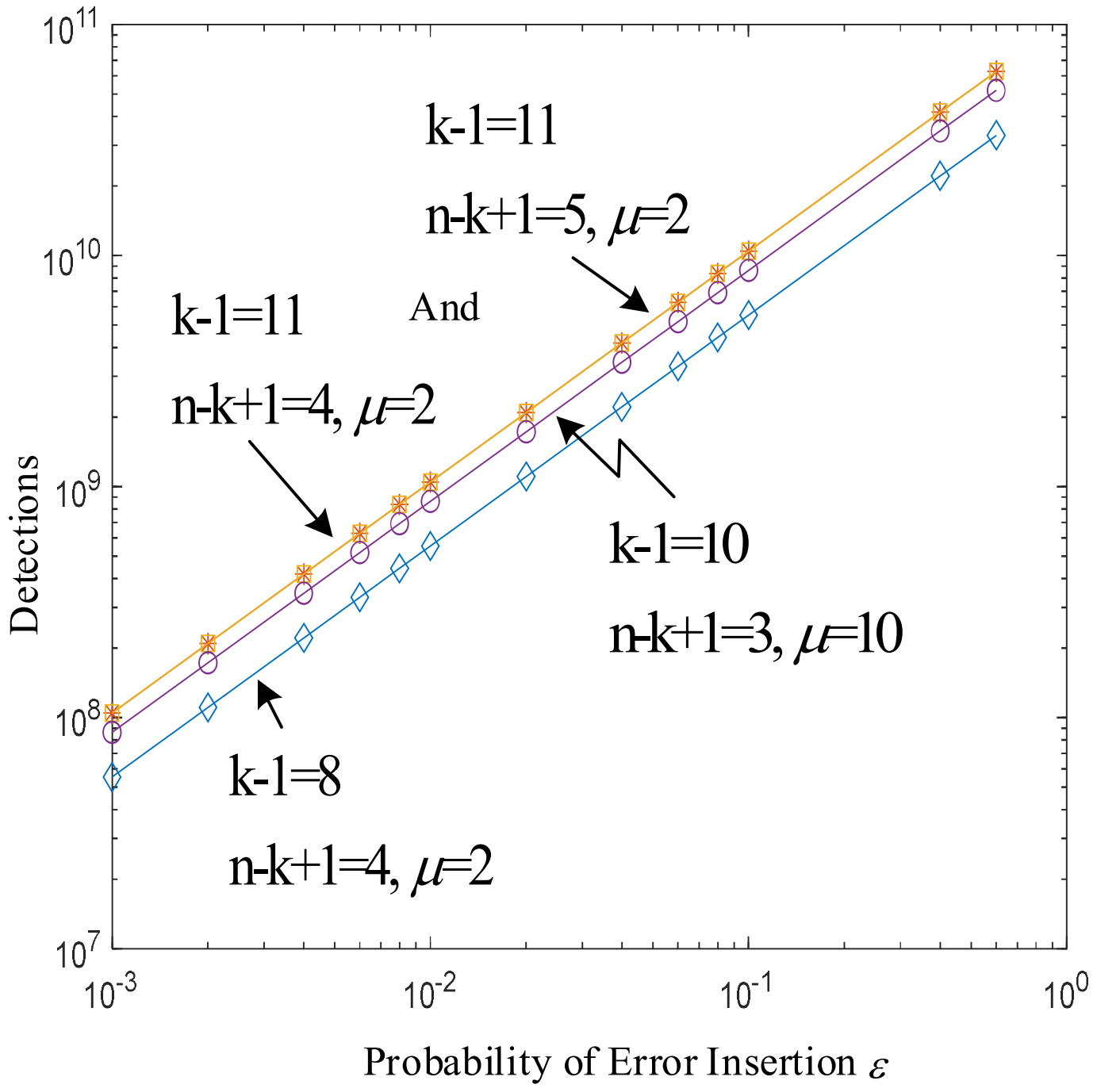

Figure 13. Detection Values of Protected Activation Functions

# A **Appendix** Codes over Number Fields

to "Safe Use of Neural Networks"

The proposed protection method for the major two parts, forward and backward processing stages, of neural network configurations involve error-detecting codes. Our codes are defined over standard arithmetic fields. We outline here more details of these types of codes, block or convolutional, so as to not interrupt the flow of the article.

### A-1 Block Codes from Binary Codes

There are wide varieties of binary error-correcting codes that can be upgraded to numerically-based codes. We will focus on the BCH class that can be constructed for large lengths, 74-150 symbols. As mentioned in a recent paper [18], real number codes can be declared by treating the binary bits as real numbers. This is a well-known result and all defining matrices now have real number 1's. [Theorem 2, 22] The error- detecting capabilities are preserved.

BCH binary codes [Sect, 6.5, 23] are a form of cyclic code that is described by a generator polynomial g(X)) whose coefficients in turn define a generator matrix. The generator polynomial of degree (n-k) for code of length n and information content k, has coefficients that also translate into a generator matrix G.

$$g(X)=g_{n-k}X^{n-k}+g_{n-k-1}X^{n-k-1}+\cdots+g_1X^1+g_0X^0 \quad \text{(A-1a)}$$

$$G=\begin{pmatrix} g_{n-k} & g_{n-k-1} & \cdots & \cdots & g_1 & g_0 & 0 & \cdots & & 0 \\ & g_{n-k} & g_{n-k-1} & & & g_1 & g_0 & 0 & \cdots & 0 \\ 0 & 0 & g_{n-k} & & & & & & \cdots & 0 \\ 0 & \cdots & 0 & & & \ddots & & & & 0 \\ & \ddots & & & & & & & & \vdots \\ \vdots & & & & & & & & & \\ 0 & \cdots & \ddots & & & & & & & \vdots \\ \vdots & & & & & & & & & 0 \\ 0 & \cdots & 0 & g_{n-k} & g_{n-k-1} & \cdots & & g_1 & g_0 & 0 \\ 0 & \cdots & \cdots & 0 & g_{n-k} & g_{n-k-1} & \cdots & \cdots & g_1 & g_0 \end{pmatrix}$$

(A-1b)

There is a difficultly in transferring the binary code to a real number code. In the binary field, -1=+1 so the binary coefficients in the original code when converted to the encoding matrix each 1 must be distinguished whether it is +1 or -1 in the real field. These distinctions can be made by examining the coefficients of g(X) when transferred.

The proper sign for the translating coefficients can be established by noting the generator polynomial is a product of first-degree polynomial factors in an extension field.

$$g(X)=\prod_{i=0}^{n-k}\left(X-b_i\right);\ b_i \text{ root in extension field} \qquad \text{(A-2)}$$

The coefficients in g(X) (A-1a) are products of the roots taken the proper number of times. For example, the confidents $g_{n-k-t}$ is a sum of all combinations of roots taken t at a time: $\sum(-b_{i_1})(-b_{i_2})\cdots(-b_{i_t})$. All items in the sum have same sign depending on the size of t, even or odd. The product will be + if t is even while product is – if t is odd. Of course, $g_{n-k}=1$, a product of all X terms. Consequently, the sign of the binary coefficients in matrix G trace back to the coefficients in g(X).

All coding features for protecting arithmetic operations should use codes in systematic form; the information symbols are completely distinguished from the easily identified parity symbols. This requires the encoding matrix to have a particular form.

$$G_S=\left(I_k \vdots P\right); I_k \text{ identity, } P\ k\times n\text{-}k \text{ parity-check matrix} \qquad \text{(A-3)}$$

With G viewed as having real arithmetic field coefficients with proper signs, it is a simple matter to apply Jordan decomposition, also called a reduced row echelon form (rref) in MatLab [26]. If $\underline{u}$ is a vector $1\times k$ holding data, the codeword associated with it is $\underline{x}$, $1\times n$, which may be partitioned into two subvectors.

$$\underline{x} = \left(\underline{u} \vdots \underline{\xi}\right); \underline{\xi} \text{ is } 1\times \text{n-k, parity vector} \qquad \text{(A-4)}$$

The parity check symbols are related to the data symbols through the submatrix P (A-3).

$$\underline{\xi} = \mathrm{P}\underline{u};\ 1\times \text{n-k, parity vector} \qquad \text{(A-5)}$$

These checking symbols are used to determine if errors have been added to the code vector $\underline{x}$.

BCH codes are defined only for certain lengths and information content due to their construction using roots in a binary extension field [23]. However, the information content, factor k, can be adjusted by setting a number of the normal information positions to 0. The information content is now (k-r) by setting r information positions to 0. The length is also shortened to (n-r). This can be described by a generator matrix like (A-1b) by selecting k-r rows of original G and removing r columns, which have nonzero components in these columns.

$$G^{(r)}=\begin{pmatrix}
g_{n-k} & g_{n-k-1} & \cdots & \cdots & g_1 & g_0 & 0 & \cdots & & 0 \\
 & g_{n-k} & g_{n-k-1} & & & g_1 & g_0 & 0 & \cdots & 0 \\
0 & 0 & g_{n-k} & & & & & & \cdots & 0 \\
0 & \cdots & 0 & & & \ddots & & & & 0 \\
 & \ddots & & & & & & & & \vdots \\
\vdots & & & & & & & & & \\
0 & \cdots & \ddots & & & & & & & \vdots \\
\vdots & & & & & & & & & 0 \\
0 & \cdots & 0 & g_{n-k} & g_{n-k-1} & \cdots & & g_1 & g_0 & 0 \\
0 & \cdots & \cdots & 0 & g_{n-k} & g_{n-k-1} & \cdots & \cdots & g_1 & g_0
\end{pmatrix} \quad \text{(k-r) rows}$$

(n-r) columns

There are (n-k+r) parity values in this code. It is possible to develop a systematic form for this shortened code using the reduced row echelon for (rref).

$$G_S^{(r)} = \left(I_{k-r} \ \vdots\ P^{(r)}\right)\ ;\ P^{(r)}\ \left((k\text{-}r)\times(n\text{-}k\text{+}r)\right)\text{parity check} \qquad \text{(A-6)}$$

The code structure may have larger detecting capabilities because of the increased number of parity values.

### A.2 Convolutional Codes

A second class of codes that can be specialized for large calculation sizes is convolutional codes. These codes allow checking procedures to be interspersed among the information data. Such codes can be described as operating on semi-infinite information streams producing checking values periodically along codeword stream. The parity values are also described by a semi-infinite checking matrix. Our convolutional codes are defined over numerical field structures. We have pioneered a class of complex-values codes based on the discrete Fourier transform pieces [24]

The codeword symbols are collected in a vector $\underline{x}$ which in turn contains a stream of subvectors $\underline{x}_i, 1\times n$.

$$\underline{x} = \left(\underline{x}_0,\ \underline{x}_1,\ \ldots,\ \underline{x}_i,\ \ldots\ldots\right)$$
$$\underline{x}_i = \left(x_{i,0},\ x_{i,1},\ \ldots,\ x_{i,n-1}\right)\ 1\times n \qquad \text{(A-7a)}$$

The codeword subvectors contain input data collected into their own subvectors that form the semi-infinite data vector $\underline{u}$.

$$\underline{u} = \left(\underline{u}_0,\ \underline{u}_1,\ \ldots,\ \underline{u}_p,\ \ldots\ldots\right)\ \text{data stream}$$
$$\underline{u}_p = \left(u_{p,0},\ u_{p,1},\ \ldots,\ u_{p,k-1}\right)\ 1\times k \qquad \text{(A-7b)}$$

While there is an encoding matrix accepting data and placing it in a codeword with parity values, the useful part for protecting data is a checking matrix H. This checking matrix annihilates all codewords $\underline{x}$, i.e., $\underline{0} = H\underline{x}$. When additive errors appear within the codeword stream, the parity-checking matrix produce a syndrome vector $\underline{S}$, which is composed of syndrome, subvectors $\underline{S}_p$.

$$\underline{S} = H\underline{x}$$
$$\underline{S}=\left(\underline{S}_0,\ \underline{S}_1\ ,\ldots,\ \underline{S}_p,\ \ldots\ldots\right);\ \underline{S}_p = \left(s_{p0},\ s_{p1},\ \ldots s_{p(n\text{-}k\text{-}1)}\right) \qquad \text{(A-8)}$$

The parity-checking matrix H is formed using a finite number of submatrices so that it engages only a finite length of the codeword stream at a time. These submatrices is involved with constraint length of the code, m.

$$H=\begin{pmatrix}
H_0 & 0 & 0 & \cdots & & & & & \\
H_1 & H_0 & 0 & 0 & \cdots & & & 0 & 0 \\
H_2 & H_1 & H_0 & 0 & 0 & \cdots & & & \\
\vdots & & \vdots & \ddots & \ddots & & & & \\
H_m & H_{m-1} & \cdots & \cdots & H_1 & H_0 & 0 & \cdots & 0 \\
0 & H_m & H_{m-1} & & & H_1 & H_0 & 0 & \cdots \\
0 & 0 & & \ddots & \ddots & & \ddots & & 0 \\
 & & & & & \ddots & & \ddots &
\end{pmatrix}$$

$$\text{each } H_i\ (n\text{-}k)\times n \qquad \text{(A-9)}$$

Each group of syndrome subvectors $\underline{S}_p$ in $\underline{S}$ involves the producting of a set of submatrices in H, collected as $H_{SEG}$.

$$H_{SEG}=\left(H_m \quad H_{m-1} \quad \cdots \quad H_1 \quad H_0\right);\ \ (n\text{-}k)\times(m\text{+}1)n \qquad \text{(A-10)}$$

This segment employs a finite memory in developing the syndromes, a characteristic of convolutional codes.

The construction of complex-values convolutional codes with parameters length n, information content k and constraint length m utilizes discrete Fourier transform vectors of length n. A basic entity in all parts of the checking matrices is an $n^{th}$ root of unity, $W=\exp\left(j2\pi/n\right)$ from which DFT vectors emerge. This class of codes, sometimes called Piret convolutional codes [24], does require a constraint among governing parameters.

$$(m+1)(n-k)<n \qquad \text{(A-11)}$$

This requirement guarantees that there are enough DFT vectors to complete the codes' constructions. A fundamental paper describing all details for constructing these types of codes is given [23] along with many other features of such codes. It is also possible to develop the generators of these DFT-based convolutional codes to have real-valued coefficients, following the fundamental policies of the original DFT codes [25].

For our use, it is best to describe these codes using a polyphase representation [26] wherein the vectors and matrices have Z-transforms of the sequences of symbols. When the components of each element in a submatrix is labeled according to its location in the submatrix, the Z-transform spreads across the submatrices as delay parameters.

$$H_r=\left(\left(h_{ij}^{(r)}\right)\right)\ ;\ \begin{matrix} i=0,1,\ldots,n\text{-}k\text{-}1 \\ j=0,1,\ldots,n\text{-}1\end{matrix} \qquad \text{(A-12a)}$$

$$h_{ij}(Z)=\sum_{r=0}^{m} h_{ij}^{(r)} Z^{-r} \qquad \text{(A-12b)}$$

$$H(Z)=\left(\left(h_{ij}(Z)\right)\right)\ ;\ \begin{matrix} i=0,1,\ldots,n\text{-}k\text{-}1 \\ j=0,1,\ldots,n\text{-}1\end{matrix} \quad n\text{-}k\times n \ \text{ matrix} \qquad \text{(A-12c)}$$

The codeword vectors and syndrome vectors also have Z-transform entities.

$$\begin{aligned} X_r(Z)&=\sum_{p=0}^{+\infty} x_{pn+r} Z^{-p} \quad ;r=0,1,\ldots(n\text{-}1) \\ S_q(Z)&=\sum_{p=0}^{+\infty} S_{p(n\text{-}k)+q} Z^{-p} \quad ;q=0,1,\ldots(n\text{-}k\text{-}1) \end{aligned} \qquad \text{(A-13)}$$

With all the notation in place, it is possible to give the syndrome equations.

$$\begin{aligned} &\underline{S}(Z)=H(Z)\underline{X}(Z) \\ &\quad \underline{S}(Z)=\left(\underline{S}_0,\ \underline{S}_1\ ,\ldots,\ \underline{S}_{n\text{-}k\text{-}1}\right);\ n\text{-}k\times 1 \\ &\quad \underline{X}(Z)=\left(\underline{X}_0,\ \underline{X}_1\ ,\ldots,\ \underline{X}_{n\text{-}k\text{-}1}\right);\ n\times 1 \end{aligned} \qquad \text{(A-14)}$$

Another possible incarnation of the code is in systematic form. There is another refinement of the code needed to get this systematic form [24]. A subcode is defined also giving still a length n code but with (k-1) information places and (n-k+1) parity values. The resulting code has parity-check equations related to polyphase quantity $H_{SUB}(Z)$.

$$H_{SUB}(Z)=\left(-\Xi(Z)\ \vdots\ I_{n\text{-}k+1}\right) \qquad \text{(A-15)}$$

$$\Xi(Z)=\sum_{i=0}^{\mu}\Xi_i Z^{-i}\ ;\ \Xi_i(n\text{-}k+1)\times(k\text{-}1)\quad \mu \text{ constraint length}$$

The syndromes have a usual equation.

$$\underline{S}_{SUB}(Z)=H_{SUB}(Z)\ \underline{X}(Z)\ ;\ (n\text{-}k+1)\times 1\ \text{ syndromes} \ \text{(A-16)}$$

On the other hand, the systematic codeword can be expressed showing two parts, data and parity.

$$\underline{X}(Z)=\begin{pmatrix}\underline{D}(Z)\\ \underline{P}(Z)\end{pmatrix}\ ;\ \begin{matrix}\underline{D}(Z)\ (k\text{-}1)\times 1\ \text{ data} \\ \underline{P}(Z)\ (n\text{-}k+1)\times 1 \text{ parity}\end{matrix} \qquad \text{(A-17)}$$

The resulting syndrome vector coming from (A-15) combined with (A-16) shows the syndromes as the sum of two items, easily recognizable.

$$\underline{S}_{SUB}(Z)=-\Xi(Z)\underline{D}(Z)+\underline{P}(Z)\quad ;(n\text{-}k+1)\times 1 \qquad \text{(A-18)}$$

This is exactly as expected observed parity subtracted from computing parity from data part.

The use of polynomial matrix transfer functions show procedures for developing the subcode's description. The systematic form (A-15) indicates that the submatrices $\Xi_i$, $(n\text{-}k+1)\times(k\text{-}1)$, $i=0,1,\ldots,\mu$ are useful in determining parity values. Expanding on the use of the convolutional code, the data components for this subcode can be collected into a semi-infinite vector containing well-defined pieces, subvecctors each (k-1) long.

$$\underline{y}=\left(\underline{y}_0,\ \underline{y}_1,\ldots,\underline{y}_i,\ \underline{y}_{i+1},\ldots\ldots\right)^T\ ;\ \underline{y}_i\ \ 1\times(k\text{-}1)\ \text{ data} \ \text{(A-19)}$$

The semi-infinite parity vector is composed of subvectors $\underline{P}_i$, $1\times(n\text{-}k+1)$, each calculated using the submatrices $\Xi_i$ shown shortly.

$$\underline{P}=\left(\underline{P}_0,\ \underline{P}_1,\ldots,\underline{P}_i,\ \underline{P}_{i+1},\ldots\ldots\right)^T\ ;\ \underline{P}_i\ \ 1\times(n\text{-}k+1)\ \text{ parity} \qquad \text{(A-20a)}$$

All subvectors result from using a parity-generating matrix $\Gamma$ with the same form as H above.

$$\Gamma=\begin{pmatrix} -\Xi_0 & 0 & 0 & \cdots & & & & & \\ -\Xi_1 & -\Xi_0 & 0 & 0 & \cdots & & & & \\ -\Xi_2 & -\Xi_1 & -\Xi_0 & 0 & 0 & \cdots & & & \\ \vdots & & \vdots & \ddots & \ddots & & & & \\ -\Xi_\mu & -\Xi_{\mu\text{-}1} & \cdots & \cdots & -\Xi_1 & -\Xi_0 & 0 & \cdots & \\ 0 & -\Xi_\mu & -\Xi_{\mu\text{-}1} & & & -\Xi_1 & -\Xi_0 & 0 & \cdots \\ 0 & 0 & & \ddots & \ddots & & \ddots & & 0 \\ & & & & & \ddots & & \ddots & \end{pmatrix} \qquad \text{(A-20b)}$$

$$\underline{P}=\Gamma\,\underline{y} \qquad \text{(A-20c)}$$

Each parity subvector $\underline{P}_i$ involves engaging a finite number of the subvectors (A-19) $\underline{y}_i$.

$$\underline{P}_q=\left(-\Xi_0,-\Xi_1,\ \cdots,-\Xi_i\right)\begin{pmatrix}\underline{y}_q\\ \underline{y}_{q+1}\\ \vdots \\ \underline{y}_{q+\mu}\end{pmatrix} \qquad \text{(A-21)}$$

$$(n\text{-}k+1)\times(\mu+1)\qquad (\mu+1)(k\text{-}1)\times 1$$

Note that $\underline{P}_{q+1}$ parity subvectors involves input subvectors $\underline{y}_{q+1}, \underline{y}_{q+2}, \cdots, \underline{y}_{q+1+\mu}$ which overlap those involve in , $\underline{P}_i$ Hence, parity symbols are generated by matrix $\Xi = \left(-\Xi_0, -\Xi_1, \cdots, -\Xi_i\right)$

The data samples in stream $\underline{y}$ can be segmented into a piece $(\mu+1)(k\text{-}1) \times 1$ and then applied to matrix $\Xi$. If there are errors in this segment of $\underline{y}$, the resulting parity values will not match those computed in a different way and method. The error detection process can proceed as segments of the observed data progress.

The ABFT technique in the text uses a modified weighting matrix combining $\Gamma$ and weighting matrix W, as $\Gamma W$. The structure of $\Gamma$ with the submatrices in $\Xi$ appearing in limited parts of $\Gamma$ clearly visible in (A-20b) means there are many fewer multiplications really in forming $\Gamma W$. There are only $(\mu+1)(k\text{-}1)$ columns nonzero when applying $\Gamma$ to W.

## AUTHOR'S BIO SKETCH

G. Robert Redinbo, who received the BS, MS, and PhD degrees in electrical engineering from Purdue University, is a Professor of Electrical and Computer Engineering at the University of California, Davis, where he was also a member of the Center for Image Processing and Integrated Computing. He has taught at Purdue University, the University of Wisconsin-Madison, where he served as a staff member, Space Sciences and Engineering Center, and at Rensselaer Polytechnic Institute, where he was a founding member of the Center for Integrated Electronics. His industrial experience includes Lockheed-Martin, the MITRE Corporation, and IBM as a visiting scientist, Data Systems Division. His US Government employment includes NASA, Defense Communications Agency, and the Army Security Agency. He has been a consultant to Sandia

National Laboratories and Thermo-Fisher. He has taken a one-year leave at the Reliability Laboratory, Swiss Federal Institute of Technology (ETH), Zurich, Switzerland, and completed a sabbatical at the Platform Components Division, Intel. He has been active in several IEEE organizations including serving as a member of the Ad Com, Acoustics, Speech, and Signal Processing Society, and as an associate editor of the IEEE Transactions on Computers. He has chaired and served on a variety of program committees for IEEE sponsored conferences and was a member of the Executive Committee, University of California MICRO Research Program. He is a Registered Professional Engineer. His research interests encompass computer engineering, fault-tolerant computing, error control in digital designs, particularly communication and signal processing. He is an extra-class amateur operator exploring low-power HF data communications employing error-correcting codes.